\documentclass[3p,times]{elsarticle}
\usepackage{framed}

\usepackage{tcolorbox}
\usepackage{fancybox} 
\tcbuselibrary{theorems}

\newtcbtheorem[number within=section]{definition}{Definition}%
{colback=gray!5,colframe=gray!10!gray,fonttitle=\bfseries}{th}

\usepackage[strict]{changepage}
\definecolor{formalshade}{gray}{0.96}
\definecolor{darkblue}{RGB}{0,80,155}
\definecolor{green}{RGB}{177,201,31}

\newenvironment{finding}{%
	\MakeFramed{\advance\hsize-\width\FrameRestore}%
	\noindent\hspace{-4.55pt}
	\begin{adjustwidth}{}{7pt}%
		\vspace{1pt}\vspace{1pt}%
	}
	{%
		\vspace{2pt}\end{adjustwidth}\endMakeFramed%
}

\usepackage{ecrc}

\volume{00}

\firstpage{1}

\journalname{Journal of Systems and Software}

\runauth{Herrmann et al.}

\jid{procs}

\jnltitlelogo{Procedia Computer Science}

\CopyrightLine{2022}{Published by Elsevier Ltd.}

\usepackage{amssymb}

\usepackage[figuresright]{rotating}

\usepackage{hyperref}

\usepackage{amsmath, amsfonts} 
\usepackage{algorithmic}
\usepackage{graphicx}
\usepackage{textcomp}
\usepackage{xcolor}

\usepackage{tikz}
\usetikzlibrary{shapes.geometric, arrows, positioning, shadows}
\usepackage{pgf-pie}
\usepackage{pgfplots}
\pgfplotsset{compat=1.17}
\usepackage{pgf}
\usepackage{colortbl}

\usepackage{enumitem}

\usepackage{tabularx}
\renewcommand{\arraystretch}{1.2}
\usepackage{makecell}
\usepackage{multirow}
\usepackage{graphicx}
\usepackage{booktabs}

\usepackage{xcolor}
\def\BibTeX{{\rm B\kern-.05em{\sc i\kern-.025em b}\kern-.08em
		T\kern-.1667em\lower.7ex\hbox{E}\kern-.125emX}}
\usepackage{framed}
\usepackage{tcolorbox}
\tcbuselibrary{theorems}

\usepackage[strict]{changepage}
\definecolor{formalshade}{gray}{0.96}
\definecolor{darkblue}{RGB}{0,80,155}
\definecolor{green}{RGB}{177,201,31}

\renewenvironment{finding}{%
	\MakeFramed{\advance\hsize-\width\FrameRestore}%
	\noindent\hspace{-4.55pt}
	\begin{adjustwidth}{}{7pt}%
		\vspace{1pt}\vspace{1pt}%
	}
	{%
		\vspace{2pt}\end{adjustwidth}\endMakeFramed%
}

\begin{document}

\begin{frontmatter}

\dochead{}

\title{On the Subjectivity of Emotions in Software Projects: How Reliable are Pre-Labeled Data Sets for Sentiment Analysis?}

\author[label1]{Marc Herrmann}
\author[label2]{Martin Obaidi}
\author[label3]{Larissa Chazette}
\author[label4]{Jil Klünder}
\address{Leibniz University Hannover\\
Software Engineering Group\\
Hannover, Germany}
\address[label1]{marc.herrmann@stud.uni-hannover.de}
\address[label2]{martin.obaidi@inf.uni-hannover.de}
\address[label3]{larissa.chazette@inf.uni-hannover.de}
\address[label4]{jil.kluender@inf.uni-hannover.de}

\begin{abstract}
Social aspects of software projects become increasingly important for research and practice. Different approaches analyze the sentiment of a development team, ranging from simply asking the team to so-called sentiment analysis on text-based communication. These sentiment analysis tools are trained using pre-labeled data sets from different sources, including GitHub and Stack Overflow.

In this paper, we investigate if the labels of the statements in the data sets coincide with the perception of potential members of a software project team. Based on an international survey, we compare the median perception of 94 participants with the pre-labeled data sets as well as every single participant's agreement with the predefined labels. Our results point to three remarkable findings: 
(1) Although the median values coincide with the predefined labels of the data sets in 62.5\% of the cases, we observe a huge difference between the single participant's ratings and the labels;
(2) there is not a single participant who totally agrees with the predefined labels; and (3) the data set whose labels are based on guidelines performs better than the ad hoc labeled data set. 
\end{abstract}

\begin{keyword}
Sentiment analysis \sep software projects \sep polarity \sep development team \sep communication


\end{keyword}

\end{frontmatter}



\section{Introduction}
The days of one-man software development are long gone~\cite{kraut1995coordination}. Today, we face software projects getting steadily bigger and more complex, requiring social efforts to work together in development teams such as adequate interaction and communication~\cite{kraut1995coordination,marjaie2011communication,mcchesney2004communication,niinimaki2012reflecting}. Inadequate and insufficient communication and interactions have been proven to be able to harm the productiveness and problem-solving abilities of developers~\cite{graziotin2014happy,graziotin2015you}. This, in turn, has a negative influence on the development team as a whole and the software project. As a result, the research field of sentiment analysis (sometimes also referred to as emotion mining) in software engineering has grown steadily over the past few years~\cite{obaidi2021development}. Nowadays, there exist multiple tools using different techniques to apply sentiment analysis to data sources from software engineering, including \textit{JIRA}~\cite{islam2017leveraging}, \textit{GitHub}~\cite{ahmed2017senticr}, and \textit{Stack Overflow}~\cite{calefato2018senti}. A recent systematic literature study~\cite{obaidi2021development} shows that almost all of these tools use some form of machine learning, requiring some kind of training data to allow the algorithm to learn which statements, comments, and texts evoke which sentiment polarity, and to be able to measure the performance of these tools. 

For this reason, there is a necessity for pre-labeled data sets to develop, train, and test such tools performing sentiment analysis in software engineering. However, manual annotation of data is very time-consuming if one wants to obtain a sufficiently large data set, and the process of labeling data is a difficult task for humans~\cite{kluender2020identifying,herrmann-kluender-2021}. One possible explanation for this difficulty is the subjectivity of the answers. Consequently, the results may not always be reliable (and they might not represent the perception of developers in general), especially, if the data points are only labeled by a single rater, even when doing so according to specified guidelines. Consequently, multiple raters per data point would be beneficial to solve this issue. Murgia et al.~\cite{Murgia.2014} state that already two independent raters labeling each data point suffice to create a reliable data set, with more than two raters per data point not having a remarkable impact on the measured degree of agreement between the raters. 

However, given the diversity of team members in software projects, including their culture, personality traits, and habits, it appears to be unlikely that gold standard data sets adequately reflect all these differences in the team. Consequently, we want to analyze how far gold standard data sets reflect the median perception of software developers as well as, more concretely, the perception of single developers. That is, would a possible team member of a software project assign the same label to a given statement as the gold standard data set? These data sets are used to train sentiment analysis tools that are meant to predict the perception of a developer or a development team (e.g., \cite{ahmed2017senticr, calefato2018senti, islam2018sentistrengthse}). Thus, in case of a data set that does not reflect the perception of a developer, it is unlikely that a tool trained on this data set adequately and correctly predicts his or her perception.

In a survey, we asked 180 participants (94 of which we considered in our final population) to rate different statements based on the perceived sentiment polarity raised when reading such statement. We did not provide any guidelines, but used excerpts from two pre-labeled data sets (from the GitHub gold standard provided by Novielli et al.~\cite{novielli2020githubgold} and an ad hoc labeled Stack Overflow data set by Lin et al.~\cite{lin18sentiment}). 

In this paper, we investigate the overall agreement between a group of raters (represented by the study participants) and the given labels, as well as the agreement between every single rater and the labels. This allows us to draw conclusions on the reliability of the pre-labeled data sets by means of which kind of team member they match (given the nature of subjective statements, it is almost impossible that the labeled data sets match every developer). 

Our results point to three interesting findings: (1) Although the median values coincide with the predefined labels of the data sets in 62.5\% of the cases, we observe a huge difference between the single participant's ratings and the labels;
(2) there is not a single participant who totally agrees with the predefined labels; and (3) the data set whose labels are based on guidelines performs better than the ad hoc labeled data set. 

\textit{Outline.} 
The rest of the paper is structured as follows: In Section~\ref{sec:background}, we present related research. Section~\ref{sec:research} presents the research design. We present our results in Section~\ref{sec:results}, which we discuss in Section~\ref{sec:discussion}. Section~\ref{sec:conclusions} concludes the paper.

\section{Background and Related Work}
\label{sec:background}

In this section, we will first explain the emotion model that was the basis for the labeling process of one of the data sets used in the remainder of this paper. Afterwards, we discuss different data sets and their annotation processes that were built in the context of software engineering. We end this section with presenting related work.

\subsection{Emotion models}
Shaver et al. \cite{shaver1987emotion} asked psychology students to perform a hierarchical sorting of emotions. This resulted in three levels in the emotion hierarchy. The first level distinguishes between positive and negative emotions. In the second level, five basic emotion categories were used: love, joy, anger, sadness, fear. The sixth emotion surprise, according to Shaver et al.~\cite{shaver1987emotion}, was not considered a basic category because it rarely occurred in the studies, but was nevertheless considered for further analysis. In the third level, emotions are divided into subcategories (for example, sadness is divided into, e.g., guilt or agony). Parrott \cite{parrott2001emotions} expanded Shaver et al.'s framework \cite{shaver1987emotion} with further details and also distinguishes between three levels: primary emotion, secondary emotion, and tertiary emotion. With the addition of surprise, the primary emotions are equal to the second level of Shaver et al.'s model~\cite{shaver1987emotion}. The tertiary emotion is a more fine-grained distinction compared to Shaver et al.~\cite{shaver1987emotion}. These models have been used to label data sets used for sentiment analysis in software engineering. 

\subsection{Sentiment analysis and data sets in software engineering}
According to a recent literature review by Obaidi and Klünder~\cite{obaidi2021development}, sentiment analysis has been in the focus of several researchers. For example, Novielli et al.~\cite{novielli2018stackgold} describe how they developed a gold standard data set to support the ongoing research on sentiment analysis in software engineering. They used annotation guidelines based on the framework by Shaver et al.~\cite{shaver1987emotion} to formalize the process of labeling statements from Stack Overflow. They considered the six emotions of the second level, without the subdivision of the third level.

The data set from Novielli et al.~\cite{novielli2018stackgold} contains 4800 data points with each data point representing a question, an answer, or a comment from Stack Overflow. Each data point was assigned a label according to the guidelines by Shaver et al.~\cite{shaver1987emotion} independently by three computer scientist student using majority voting to receive the final sentiment label for the data set. The Fleiss' $\kappa$~\cite{fleiss1971agreement} values ranged from 0.30 to 0.66. However, since the observed agreement (i.e., the percentage of cases for which raters provided the same annotation) ranged from 0.87 to 0.98, they consider their data set being reliable. The rather small values of Fleiss' $\kappa$ can be explained by the remarkable chance of random agreement~\cite{novielli2018stackgold}.

In a follow-up paper, Novielli et al.~\cite{novielli2020githubgold} created an even larger gold standard data set from GitHub pull-request and commit comments. They used the same methodology and guidelines by Shaver et al.~\cite{shaver1987emotion} with three raters independently assigning labels to each data point. This time, Novielli et al.~\cite{novielli2020githubgold} mapped the affects from the framework~\cite{shaver1987emotion} into the three sentiment polarity classes with love and joy being \textit{positive}, and sadness, anger, and fear being \textit{negative}. The ``surprise'' category was discarded because this emotion could not be transferred adequately~\cite{novielli2020githubgold}. All other data points retrieved the class \textit{neutral} because of the absence of explicit emotions. In total, the authors obtained 7113 data points classified into \textit{positive}, \textit{negative} , and \textit{neutral}~\cite{novielli2020githubgold}. This is one of the largest pre-labeled data set for sentiment analysis in the software engineering domain that is published\footnote{\href{https://figshare.com/ndownloader/files/21001260}{The GitHub gold standard data set is available on Figshare.}}.

Murgia et al. \cite{Murgia.2014}, analogously to Novielli et al. \cite{novielli2020githubgold,novielli2018stackgold,}, annotated a data set consisting of JIRA issue comments using an emotion model. They used the six primary emotions of level one from Parrott's framework \cite{parrott2001emotions}, which consists of the framework by Shaver et al. \cite{shaver1987emotion} as mentioned before. Among other things, they found that more than two raters do not seem to make a significant difference in terms of agreement. For this they compared Cohen's $\kappa$~\cite{cohen1960agreement} for two raters with Fleiss' $\kappa$~\cite{fleiss1971agreement} for more than 2 raters.

Lin et al.~\cite{lin18sentiment} analyze the limits of sentiment analysis in the software engineering domain. For this purpose, they manually labeled sentences extracted from Stack Overflow. As there is no mention of the use of some kind of guidelines or framework to assign the sentiment polarity labels to the data set, we presume that ad hoc labeling was used. Zhang et al.~\cite{zhang20sentiment} use the data set by Lin et al.~\cite{lin18sentiment} in their replication study. They tested various software engineering specific sentiment analysis tools and pre-trained transformer-based models on six different data sets from the software engineering domain~\cite{zhang20sentiment}. 

Ahmed et al. \cite{ahmed2017senticr} build a data set, which includes negative and non-negative documents. Therefore it is a binary-class data set. The documents consists of multiple comments. They collected them from 20 open-source projects that practice tool-based code reviews supported by the same tool (e.g., Gerrit). Three of the authors labeled the sentences based on their perception as recipients of the message. Thus, the evaluation was probably ad hoc.

However, all these papers stated that after discussing or majority voting, they resolved the disagreement cases and thus built a data set for evaluation, without having these data sets validated by external computer scientists.

\subsection{Application of sentiment analysis in software engineering}

Obaidi and Klünder~\cite{obaidi2021development} examined the development and application of sentiment analysis in software engineering through a systematic literature review ($n$ = 80 papers). Among other results, their findings show that both GitHub and Stack Overflow are among the top three data sources used for sentiment analysis. In addition, domain independent sentiment analysis tools have been found to often leading to poor performance in the software engineering domain, because of certain terms being used differently in software engineering than elsewhere (e.g., ''to kill a process")~\cite{obaidi2021development}. Those two findings reinforce the value of the data sets by Novielli et al.~\cite{novielli2020githubgold} and Lin et al.~\cite{lin18sentiment}. However, Obaidi and Klünder~\cite{obaidi2021development} also found that a common difficulty for authors lies in the subjectivity of labeling a statement with a sentiment alone, meaning that different people would already assign different labels to the same statements~\cite{zhang20sentiment,imtiaz18sentiment, Ding.2018, 8643972}.
Imtiaz et al. \cite{imtiaz18sentiment} for example annotated GitHub comments based on the annotation guide provided by Mohmmad \cite{mohammad2016practical}. A total of 3 coders were involved in the sentiment annotation process, but their Cohen's $\kappa$ values ranged from 0.27 to 0.38. They suggested that more coders could ensure more reliable human evaluation.

This commonly faced issue motivated the research on the subjectivity of sentiment perception in this paper. 

In this paper, we compare the sentiment labels assigned by different raters from the gold standard data set by Novielli et al.~\cite{novielli2020githubgold} and the ad hoc annotated data set from Lin et al.~\cite{lin18sentiment} with the  perception of possible software project team members (without using any guidelines) in the software engineering domain. To achieve this, we analyzed labeling data from 94 participants in a survey and compare the participants' results to the excerpts of two publicly available pre-labeled data sets~\cite{novielli2020githubgold, lin18sentiment}.

\section{Survey Design}
\label{sec:research}

In the following subsections, we present our research objective and the research questions, as well as the structure and our ideas behind the creation of the survey. 

\subsection{Research Objective and Research Questions}
Our main research objective is to compare the perceptions of software project team members with the polarity provided by pre-labeled data sets for sentiment analysis. This helps improve the reliability and accuracy of predictions for sentiment analysis tools in software engineering domains. To achieve this goal, we pose the following research questions:

\begin{itemize}[leftmargin=.38in]
    \item[RQ1:] How do the median labels assigned by the study participants differ from the predefined labels? 
    \item[RQ2:] How much do each participant's labels differ from the predefined labels?
    \item[RQ3:] How do the results differ between ad hoc and guideline-based labeled data sets?
\end{itemize}

\subsection{Instrument Development}
We used the survey method~\cite{robson2016real} to collect our data, implemented as an online questionnaire. 

\subsubsection{Survey Structure}
The final questionnaire consisted of five blocks of questions (number of questions in parentheses): Demographics (4), affiliation to computer science (2), programming experience (5), labeling (1 repeated for 100 statements), criteria (1). A detailed overview of the survey structure is presented in Table~\ref{table:survey}. 

\begingroup
\renewcommand*{\arraystretch}{1.35}
\begin{table}[htbp]
\caption{Survey design}

\begin{center}
\resizebox{0.5\columnwidth}{!}{%
\begin{tabular}{|c|c|c|}
\cline{2-3}
\multicolumn{1}{c|}{} & {Questions} & {Answer Options} \\
\hline

\multirow{10}{*}{\rotatebox{90}{{Demographics}}} 
 & How old are you? & 18 - 99 \\ \cline{2-3}
 & \multirow{2}{*}{What is your gender?}
  & Male \\ \cline{3-3}
  & \multicolumn{1}{c|}{} & Female \\ \cline{2-3}
 & \multirow{2}{*}{Is English your native language?}
  & Yes \\ \cline{3-3}
  & \multicolumn{1}{c|}{} & No \\ \cline{2-3}
 & \multirow{5}{*}{\makecell{How often do you\\communicate in English?}} 
  & Every day \\ \cline{3-3}
  & \multicolumn{1}{c|}{} & Multiple times a week \\ \cline{3-3}
  & \multicolumn{1}{c|}{} & Once a week \\ \cline{3-3}
  & \multicolumn{1}{c|}{} & Once in a while \\ \cline{3-3}
  & \multicolumn{1}{c|}{} & Never \\ 
  \hline
  
\multirow{8}{*}{\rotatebox{90}{{Affiliation to CS}}} 
 & \multirow{2}{*}{\makecell{Would you identify yourself\\as computer scientist?}}
  & Yes \\ \cline{3-3}
  & \multicolumn{1}{c|}{} & No \\ \cline{2-3}
 & \multirow{6}{*}{What is your professional status?}
  & Student \\ \cline{3-3}
  & \multicolumn{1}{c|}{} & Working in Academia \\ \cline{3-3}
  & \multicolumn{1}{c|}{} & Working in Industry \\ \cline{3-3}
  & \multicolumn{1}{c|}{} & Retired \\ \cline{3-3}
  & \multicolumn{1}{c|}{} & Unemployed \\ \cline{3-3}
  & \multicolumn{1}{c|}{} & Other \\ \cline{2-3}
  \hline
  
\multirow{9}{*}{\rotatebox{90}{{Programming Experience}}}
 & \multirow{2}{*}{\makecell{Do you have any experience\\with programming?}}
  & Yes \\ \cline{3-3}
  & \multicolumn{1}{c|}{} & No \\ \cline{2-3}
 & \makecell{How would you rate your\\programming skills?} & (bad) 1 - 5 (good)\\ \cline{2-3}
 & \makecell{How many years of professional\\experience do you have\\as a developer?} & 0 - 99 \\ \cline{2-3}
 & \makecell{How familiar are you with\\working as a developer\\with a team?} & (hardly) 1 - 5 (highly)\\ \cline{2-3}
 & \makecell{How many years of professional\\experience do you have in\\working with a team?} & 0 - 99 \\ \cline{2-3}
 \hline
 
\multirow{3}{*}{\rotatebox{90}{{Labeling}}} 
 & \multirow{3}{*}{\makecell{How would you label the following\\sentences regarding its polarity\\based on your perception?\\\scriptsize(for 100 statements)}}
  & Positive \\ \cline{3-3}
  & \multicolumn{1}{c|}{} & Neutral \\ \cline{3-3}
  & \multicolumn{1}{c|}{} & Negative \\ \cline{3-3}
  \hline
  
\multirow{3}{*}{\rotatebox{90}{{Criteria}}} 
 & \multirow{3}{*}{\makecell{What criteria did you use to\\decide on the polarities\\of those sentences?}}
  & Content \\ \cline{3-3}
  & \multicolumn{1}{c|}{} & Tone \\ \cline{3-3}
  & \multicolumn{1}{c|}{} & Other \\ \cline{3-3}
  \hline


\end{tabular}%
} 
\label{table:survey}
\end{center}
\end{table}
\endgroup

In the beginning, we collected demographic data such as age, gender, if English is the native language, and the frequency of communication on English. The next block of questions asked whether the participants identify as computer scientists, and what their current professional status is. Next, the questions dealt with the programming skills, the experience in professional work environments and with team work. 

Thereafter, the survey had a block of 100 statements. The 100 statements consisted of 48 different statements from the gold standard data set~\cite{novielli2020githubgold}, and 48 statements from the ad hoc labeled data set~\cite{lin18sentiment}, and 4 duplicates. We added 4 random statements (2 \textit{positive}, 1 \textit{neutral}, and 1 \textit{negative}) from the GitHub gold standard data set~\cite{novielli2020githubgold} twice to make total of 100 statements. This allows to check how consistent the participants labeled the statement, i.e. to recognize if they choose two different labels for any of the duplicate statements during the survey. The participants were asked to classify each statement as one of the three polarity classes \textit{positive}, \textit{neutral}, and \textit{negative} (without any given guidelines). The selection of these statements is presented in more detail in the next subsection. The 100 statements in the survey were randomly selected in a way leading to 32 \textit{positive}, 32 \textit{neutral}, and 32 \textit{negative} statements (plus 4 duplicates), making a nearly perfect one third split for each of the three sentiment polarity classes\footnote{Note that we did not tell the participants that the statements are equally distributed among the polarity classes to avoid biases.}. The statements were put together in blocks of ten statements. Both the ten blocks and the ten statements in each block were randomly ordered. 

The last question asked the participants to explain how they selected the labels by presenting predefined answer options such as having focused on the statement content or the statement tone, as well as a free-text answer.

\subsubsection{Selection of Reference Data Sets}
Answering the research questions requires a comparison between the data provided by our survey and already existing data sets. Consequently, we wanted the participants to label statements from data sets that are established when training sentiment analysis tools in software engineering. We selected two different data sets (and thereby two different platform sources) with one being guideline-based annotated and one being ad hoc labeled. The GitHub gold standard data set from Novielli et al.~\cite{novielli2020githubgold} was selected as guideline-based data set and the Stack Overflow data set from Lin et al.~\cite{lin18sentiment} was the reference for the ad hoc labeled data set. The GitHub data set~\cite{novielli2020githubgold} is the largest data set (7122 statements) to the best of our knowledge and was built most recently (in 2020). It was used and evaluated in the context of sentiment analysis in software engineering in previous papers (e.g., \cite{novielli2020githubgold, zhang20sentiment, bertcomparison2021}). The Stack Overflow data set~\cite{lin18sentiment} contains 1500 statements and was used also in many papers (e.g., \cite{lin18sentiment, zhang20sentiment,emoji2022}).

From each of the two data sets, we randomly selected 48 statements with 16 \textit{positive}, 16 \textit{negative}, and 16 \textit{neutral} statements, resulting in a total of 96 statements. We evaluate the performance of both data sets separately to compare both guideline-based and ad-hoc annotation to the perception of the participants, as well as both data sets in combination to evaluate the general discrepancies between the labels predefined by the scientific authors and the participants. Please note that, for simplicity, we will refer to these 48 selected statements as the Stack Overflow and GitHub data sets as well as the combined data set (96 statements) in the remainder of the paper.

\subsection{Data Collection}
The survey was hosted using \textit{LimeSurvey} on the server of the Software Engineering Group at Leibniz University Hannover. 
We mainly distributed the survey via e-mail. We invited computer science students (BSc and MSc) at Leibniz University Hannover attending our lectures\footnote{Note that, due to privacy reasons, we were not allowed to invite our students (or students from other universities) via e-mail, but we distributed the information on our survey via the internal software tool used to support lectures. This way, we only contacted students from Leibniz University Hannover, but we asked colleagues from other universities to do the same and to share the survey with their students.}, as well as computer science doctoral students, postdocs, and research assistants (for which we manually extracted the contact information from the institutional websites).
In addition, the survey was sent to publicly available e-mail addresses from researchers who published papers about sentiment analysis in software engineering themselves to ask them to participate and to share the survey with their networks. We extracted this list of authors from a recent systematic literature study~\cite{obaidi2021development}. We also shared the survey via social networks such as Twitter, Facebook, LinkedIn, and XING. In the invitation text, we included a short description of the survey, the estimated time of 10 - 20 minutes, and the link to the survey. When distributing the survey, we clearly stated that our target group are computer scientists with programming experience. However, to mitigate the risk of retrieving answers from outside the target group, we inserted three questions asking for (1) the identification as a computer scientist or developer, (2) programming experience, and (3) experiences with software development in teams. The survey was available from the end of April 2021 until the beginning of November 2021. The raw data set is available via \href{https://doi.org/10.5281/zenodo.6611728}{Zenodo}~\cite{sentisurvey-zenodo}.

\subsection{Data Pre-Processing}
\label{subsec:preprocess}
In total, we received 180 responses. We collected 127 parameters for each data point, some of them resulting from optional questions. 

The definition of a subset of the 180 data points being suitable for the analysis presented in this paper was done based on two conditions. As an incomplete data point can still provide interesting insights for answering the research questions, we did not exclude incomplete data points per s\'e. First, we removed 17 data points that answered either question on programming experience or being a computer scientist (\textit{``Would you identify yourself as a computer scientist (e.g., computer scientist student, developer, etc.)?\hphantom{}''} or  \textit{``Do you have any experience with programming (e.g programming a software, website, an app, etc.)?\hphantom{}''}) with a ``No'' as our target group were persons who are potential team members in a software project team. As a second criterion, we only included data points where participants had at least annotated one of the statements with their perceived sentiment polarity, removing additional 69 data points. This decision leads to a varying $n$-value for the single statements for the different analysis steps. Thus, we report the $n$-value separately for the different analyses. 
Applying these criteria led to a final data set consisting of 94 data points that were used for the further analyses. 

\subsection{Data Analysis}
\label{subsec:analysis}
We analyzed our collected and pre-processed data using Python with various packages for scientific research including \textit{pandas}~\cite{mckinney2010pandas}, \textit{NumPy}~\cite{harris2020numpy}, \textit{Matplotlib}~\cite{hunter2007matplotlib}, and \textit{scikit-learn}~\cite{pedregosa2011scikit}. 


\subsubsection{Analysis Procedures for RQ1} To answer our first research question, we calculated the median sentiment class from all the participants' annotations for each statement using the ordinal order \textit{negative} $<$ \textit{neutral} $<$ \textit{positive}. Note that we opted for the median rather than for a majority vote (which is often used incorrectly in the context of sentiment analysis) for the following reason: Assume we have three raters/voters and each of them assigns a different sentiment polarity class to a given statement. The votes are therefore \textit{negative}, \textit{neutral}, and \textit{positive}. The majority vote is ambiguous in this case, as each class received the same amount of votes, and a random sentiment polarity class would be returned with common methods. The median is well defined on the other hand, because we take the ordinal order \textit{negative} $<$ \textit{neutral} $<$ \textit{positive} into account. The sentiment polarity class \textit{neutral} has an even amount of sentiment polarity values above and below it (one each). For this reason we have decided to use the median instead of majority vote for our analysis. We then compared the resulting median sentiment labels with the predefined labels from the ad hoc labeled Stack Overflow data set, the guideline-based labeled GitHub data set, and both data sets in combination. We calculated the absolute and relative counts of coinciding labels (median label vs. predefined label) as well as cases of mild and severe disagreement, as proposed by \cite{novielli2020githubgold}. Finally, we calculated the observer agreement using Cohen's $\kappa$~\cite{cohen1960agreement}. That is, we considered the following four metrics:

\begin{itemize}
    \item \textbf{Agreement (also referred to as perfect agreement):} Absolute and relative amount of statements that are annotated with the same label both in the original data sets and from the participant(s)
    \item \textbf{Mild Disagreement:} Absolute and relative amount of statements that are annotated with differentiating labels in the original data sets and from the participant(s) with a distance of 1 (i.e. positive - neutral, neutral - negative)
    \item \textbf{Severe Disagreement:} Absolute and relative amount of statements that are annotated with differentiating labels in the original data sets and from the participant(s) with a distance of 2 (i.e. positive - negative)
    \item \textbf{Cohen's $\kappa$:} Calculation of the interrater agreement between the predefined labels and the participant(s)
\end{itemize}

For a more detailed description of the differences between the participants' median labels and the predefined labels, we also calculated precision, recall and $F_1$-score for each of the three data sets (from Stack Overflow, from GitHub, and in combination), as well as the confusion matrices. To calculate these values, we assumed the calculated median labels for each statement as the predicted label and used the predefined labels as the true labels.

\subsubsection{Analysis Procedures for RQ2}
To answer our second research question, we explored our data in more detail. That is, we calculated the agreement and Cohen's $\kappa$~\cite{cohen1960agreement} between each set of individual participant labels and the predefined labels from the two data sets and the combined data set. Note that we only calculated the values if the individual participant had annotated at least 10\% of statements from the corresponding data set, as we assumed an amount of less then 10\% leading to too few data points to compare so that the results would not be meaningful. The six resulting arrays (agreement and Cohen's $\kappa$ for the three data sets (from Stack Overflow, from GitHub, and in combination) of agreement and Cohen's $\kappa$ were analyzed descriptively by calculating minimum, maximum, mean, and standard deviation values. 

\subsubsection{Analysis Procedures for RQ3}
To answer RQ3, we compared the results of the analyses for RQ1 and RQ2. To measure if the difference of participant agreement and $\kappa$-values between the two data sets is statistically significant, we applied hypotheses testing. In particular, we tested the main hypothesis H1 presented in Table~\ref{table:hypotheses} at a significance level of $p < 0.05$. However, as the main hypothesis asks for a difference between the results for the two data sets in general, we formulated the two sub-hypotheses H1.1 and H1.2. To analyze H1.1 and H1.2, we first tested the respective data for normal distribution using the Shapiro-Wilk test~\cite{shapiro1965}. In case of normal distribution, we used the repeated-measures t-test~\cite{student1908} to test the hypothesis. Otherwise, we opted for the Wilcoxon signed-rank test~\cite{rey2011wilcoxon}.  As we tested two sub-hypotheses for one main hypothesis H1, we applied the Bonferroni correction~\cite{haynes2013} leading to an adjusted significance level of $p_{corr} = p/2 = 0.025$. As soon as one of the sub-hypotheses results in a $p$-value lower than $p_{corr}$, we consider H1 to be significant. Note that we only considered data points of participants who have labeled at least 10\% of statements in both of the two data sets.

\begin{table}[ht]
	\caption{Null hypotheses to compare the two data sets}
    \begin{center}
	\begin{tabular}{@{}p{0.085\columnwidth}p{0.8\columnwidth}@{}}
		\Xhline{2\arrayrulewidth}
		H1$_0$ & There is no difference between the two data sets by means of the predefined and the participant's labels. \\
		H1.1$_0$ & There is no difference in the agreement between the two data sets comparing the predefined labels and the participant's labels. \\
		H1.2$_0$ & There is no difference in Cohen's $\kappa$ between the two data sets comparing the predefined labels and the participant's labels. \\
		\Xhline{2\arrayrulewidth}
	\end{tabular} 
	\end{center}
	\label{table:hypotheses}
\end{table}

\section{Results}
\label{sec:results}
We conducted the data cleaning and the data analysis as described in Section~\ref{subsec:preprocess} and Section~\ref{subsec:analysis}. In the following subsections, we present the results of each analysis step. 

\subsection{Demographics}
The participants had an average age age of 27.61 years (min: 18 years, max: 55 years, SD: 6.57 years, $n$ = 93). Seventy-seven~(81.9\%) of the participants stated their gender as male and 17~(18.1\%) as female ($n$ = 94). Only 2~(2.2\%) participants were native English speakers, and 91~(97.8\%) were foreign language speakers ($n$ = 93). However, only 2~(2.2\%) participants reported that they never communicate in English (but still considered their English as sufficient to participate in the study), while 30~(33.0\%) participants reported on communicating in English once in a while, 13~(14.3\%) participants once a week, 20~(22.0\%) participants multiple times a week, and 26~(28.6\%) participants every day ($n$ = 91). For their professional status most of the participants reported being students (66 participants; 70.2\%), while 27~(28.7\%) participants were working in industry, 15~(16\%) participants were working in academia, 3~(3.2\%) participants selected the category ''Other``, and 2~(2.1\%) participants were unemployed ($n$ = 94, multiple answers were possible). The participants rated their own programming skills on a Likert scale from 1 (beginner) to 5 (expert) with a median of 3 ($n$ = 94). On average, participants had 2.957 years of professional experience as a developer (min: 0 years, max: 25 years, SD: 4.674 years, $n$ = 93). Considering the familiarity of working as a developer within a team (Likert scale from 1 (no experience) to 5 (very experienced)) the participants stated a median value of 3 ($n$ = 93). The years of professional experience as a developer working in a team ranged from 0 to 25 years with an average of 1.957 years (SD: 3.736 years, $n$ = 92).

\subsection{Comparison Between Median Labels and Predefined Labels}
The overall results of agreement, the matches between the median participants perceptions and the predefined labels, and the resulting value of Cohen's $\kappa$ are summarized in Table~\ref{table:rq1.1}. Note that, in this step, we compare the median rating of all participants in the study with the predefined labels of the data set. In the following, we present more detailed results on the whole data set and distinguish between the statements originating from Stack Overflow and GitHub. 

\begin{table}[htbp]
	\caption{Comparison between median participants' and predefined labels}
	\begin{center}
		\begin{tabular}{@{}ll@{\hskip 11mm}llll@{}}
		    \Xhline{2\arrayrulewidth}
		    \addlinespace
			Data set & \rotatebox{90}{Size} & \rotatebox{90}{Agreement} & \rotatebox{90}{Mild Disagr.} & \rotatebox{90}{Severe Disagr.} & \rotatebox{90}{Cohen's $\kappa$} \\
			\hline
			Combined & 96   & 60 (62.5\%) & 29 (30.2\%) & 7 (7.7\%) & 0.438 \\  
			\hline
			SO & 48   & 27 (56.3\%) & 21(43.7\%) & 0 (0.0\%) & 0.344\\
			\hline
			GitHub & 48   & 33 (68.8\%) & 8 (16.7\%) & 7 (14.6\%) & 0.531\\
			\Xhline{2\arrayrulewidth}
		\end{tabular} 
		\label{table:rq1.1}
	\end{center}
\end{table}

\subsubsection{Overall Agreement}
In total, in 60 out of 96 cases (62.5\%) the median rating of the participants coincides with the predefined labels in the two data sets~\cite{novielli2020githubgold,lin18sentiment}. This agreement results in a Cohen's $\kappa$-value of 0.4375, which can be considered moderate agreement according to the scale provided by Landis and Koch ~\cite{landis1977measurement}.

In case of the ad hoc labeled Stack Overflow data set, for 27 out of 48 statements (56.25\%) the median of our participants coincides with the labels given in the data set. This is considered fair agreement with a Cohen's $\kappa$ of 0.34375. In the 21 (43.7\%) cases where the median label of the participants does not coincide with the labels given by the data set the disagreement is \textit{always} mild, i.e. there are no cases of severe disagreement.


For the 48 statements from the GitHub data set, we retrieve an agreement between our participants and the predefined labels of 68.75\% (33 out of 48 statements are assigned the same label). This results in a Cohen's $\kappa$ of 0.53125, which is considered moderate agreement. However, there are 7 (14.6\%) cases of severe disagreement. Additionally 8 (16.7\%) cases of mild disagreement between the median labels of the participants and the predefined labels occurred. 


\begin{finding}
    \textbf{Finding 1:} For the 96 statements, we find a moderate agreement between the participants perceptions and the predefined labels with a Cohen's $\kappa$ of 0.4375.\\
    \textbf{Finding 2:} For the 48 ad hoc pre-labeled statements, there is a fair agreement ($\kappa = 0.3438$) between the participants and the predefined labels and no cases of severe disagreement.\\
    \textbf{Finding 3:} For the 48 guideline-based labeled statements, the agreement between the participants and the predefined labels is moderate ($\kappa = 0.5313$), but there are 7 cases of severe disagreement (14.6\%).
\end{finding}

\subsubsection{Precision, Recall, and F1}
Looking into more details of the (dis-)agreements between the median values of our participants and the predefined labels of the data sets leads to the results presented in Table~\ref{table:rq1.2}. The table summarizes the precision, recall, and $F_1$-score for each data set and each sentiment polarity class. 

\begin{table}[ht]
	\caption{Precision (P), recall (R), and $F_1$-score for each data set}
	\begin{center}
		\begin{tabular}{@{}cllll@{}}
			\Xhline{2\arrayrulewidth}
			\multicolumn{1}{c}{} & Polarity Class & {P} & {R} & {F1} \\ 
			\hline
			\multirow{3}{*}{\makecell{\textit{Combined}\\\textit{data set}}}
			& Positive & 0.770 & 0.313 & 0.444 \\
			& Neutral & 0.527 & 0.906 & 0.666 \\
			& Negative & 0.750 & 0.656 & 0.700 \\  
			\hline
			\multirow{3}{*}{\makecell{\textit{Stack Overflow}\\\textit{data set (ad hoc)}}}
			& Positive & 0.667 & 0.250 & 0.364\\
			& Neutral & 0.424 & 0.875 & 0.571\\
			& Negative & 1.00 & 0.563 & 0.720 \\  
			\hline
			\multirow{3}{*}{\makecell{\textit{GitHub data set}\\\textit{(guideline-based)}}}
			& Positive & 0.857 & 0.375 & 0.522 \\
			& Neutral & 0.682 & 0.938 & 0.789 \\
			& Negative & 0.632 & 0.750 & 0.686 \\  
			\Xhline{2\arrayrulewidth}
		\end{tabular} 
		\label{table:rq1.2}
	\end{center}
\end{table}

For the whole data set (i.e., the combination of statements from both data sets), we have the best precision (0.77) for the \textit{positive} class, which has the worst recall (0.313), resulting in a F1-score of 0.444. The negative class reaches almost the same precision (0.75) and a recall of 0.656. The neutral class has a precision of 0.527 and achieves the highest recall value of 0.906. 

The Stack Overflow data set achieves a perfect precision of 1.00 for the negative class, meaning that all negative values assigned by the participants are predefined to be negative by the data set. However, as the participants did not ``find'' all negative statements (but considered some as neutral), the recall of the negative class is 0.563. The positive class achieves a precision of 0.667 and the lowest recall of 0.25, meaning that several positive statements have not been identified as positive. The neutral class has the lowest precision with 0.424 and the highest recall of 0.875, meaning that several statements classified as neutral are positive or negative, and that several neutral statements have been identified as neutral.

The GitHub data set has in total the highest values for precision, recall, and F1-score. The positive class has again the highest precision (0.857), but the lowest recall (0.375). The negative class has a precision of 0.632 and a recall of 0.750. The neutral class has a precision of 0.682 and a recall of 0.938. 

\begin{finding}
    \textbf{Finding 4:} The neutral class always has the highest recall, but the worst precision. Thus, our participants were good at identifying the predefined neutral labels as neutral, but in addition classified many other statements as neutral although they had a predefined label of positive or negative. \\
    \textbf{Finding 5:} The positive class often has the lowest recall, but the best precision (except for SO). Thus, most statements labeled as positive by our participants were also predefined to be positive, but a lot of predefined positive statement were instead assigned labels of either neutral or negative by the participants.
\end{finding}

\subsubsection{Confusion Matrices}
The confusion matrices for the whole data set and the predefined labels in the data sets from Stack Overflow and GitHub are shown in Table~\ref{table:confusionmatrix}.

\begingroup
\renewcommand*{\arraystretch}{1.5}
\begin{table}[tbp]
	\caption{Confusion matrices for the data sets}
	
	\definecolor{heatmap16}{HTML}{000000}
	\definecolor{heatmap15}{HTML}{141414}
	\definecolor{heatmap14}{HTML}{212121}
	\definecolor{heatmap13}{HTML}{2e2e2e}
	\definecolor{heatmap12}{HTML}{3b3b3b}
	\definecolor{heatmap11}{HTML}{4a4a4a}
	\definecolor{heatmap10}{HTML}{585858}
	\definecolor{heatmap09}{HTML}{676767}
	\definecolor{heatmap08}{HTML}{777777}
	\definecolor{heatmap07}{HTML}{878787}
	\definecolor{heatmap06}{HTML}{979797}
	\definecolor{heatmap05}{HTML}{a8a8a8}
	\definecolor{heatmap04}{HTML}{b9b9b9}
	\definecolor{heatmap03}{HTML}{cacaca}
	\definecolor{heatmap02}{HTML}{dbdbdb}
	\definecolor{heatmap01}{HTML}{ededed}
	\definecolor{heatmap00}{HTML}{ffffff}
	
	\begin{center}
		\begin{tabular}{ccccc}
		    \Xhline{2\arrayrulewidth}
		    \multicolumn{1}{c}{} &\multicolumn{1}{c}{} &\multicolumn{3}{c}{Median Label} \\ \cline{3-5} 
			\multicolumn{2}{c}{Predefined Label} & 
			\multicolumn{1}{c}{Positive} &
			\multicolumn{1}{c}{Neutral} &
			\multicolumn{1}{c}{Negative} \\ 
			\hline
			\multirow{3}{*}{\makecell{\textit{Combined}\\\textit{data set}}}
			& Positive & \cellcolor{heatmap05}10 & \cellcolor{heatmap08}\color[HTML]{FFFFFF}{16} & \cellcolor{heatmap03}6 \\ 
			& Neutral  & \cellcolor{heatmap01}2 & \cellcolor[HTML]{1A1A1A}\color[HTML]{FFFFFF}{29} & \cellcolor[HTML]{F6F6F6}1 \\ 
			& Negative & \cellcolor[HTML]{F6F6F6}1 & \cellcolor{heatmap05}10 & \cellcolor[HTML]{585858}\color[HTML]{FFFFFF}21  \\ 
			\hline
			\multirow{3}{*}{\makecell{\textit{Stack Overflow}\\\textit{data set (ad hoc)}}}
			& Positive & \cellcolor{heatmap04}4 & \cellcolor{heatmap12}\color[HTML]{FFFFFF}{12} & 0 \\ 
			& Neutral  & \cellcolor{heatmap02}2 & \cellcolor{heatmap14}\color[HTML]{FFFFFF}{14} & 0 \\ 
			& Negative & 0 & \cellcolor{heatmap07}\color[HTML]{FFFFFF}7 & \cellcolor{heatmap09}\color[HTML]{FFFFFF}{9}  \\
			\hline
			\multirow{3}{*}{\makecell{\textit{GitHub data set}\\\textit{(guideline-based)}}}
			& Positive & \cellcolor{heatmap06}{6} & \cellcolor{heatmap04}4 & \cellcolor{heatmap06}{6} \\ 
			& Neutral  & 0 & \cellcolor{heatmap15}\color[HTML]{FFFFFF}{15} & \cellcolor{heatmap01}1 \\ 
			& Negative & \cellcolor{heatmap01}1 & \cellcolor{heatmap03}3 & \cellcolor{heatmap12}\color[HTML]{FFFFFF}{12}  \\
			\Xhline{2\arrayrulewidth}
		\end{tabular}
		
		\vspace*{1.5mm}
		
		\ifx
		\resizebox{0.825\columnwidth}{!}{%
			\begin{tabular}{cccccccccccccccc}
				\hline
				\cellcolor{heatmap00}0 & \cellcolor{heatmap01}1 & \cellcolor{heatmap02}2 & \cellcolor{heatmap03}3 &
				\cellcolor{heatmap04}4 & \cellcolor{heatmap05}5 & \cellcolor{heatmap06}6 & \cellcolor{heatmap07}\color[HTML]{FFFFFF}7 &
				\cellcolor{heatmap08}\color[HTML]{FFFFFF}8 & \cellcolor{heatmap09}\color[HTML]{FFFFFF}9 & \cellcolor{heatmap10}\color[HTML]{FFFFFF}10 &
				\cellcolor{heatmap11}\color[HTML]{FFFFFF}11 & \cellcolor{heatmap12}\color[HTML]{FFFFFF}12 & \cellcolor{heatmap13}\color[HTML]{FFFFFF}13 &
				\cellcolor{heatmap14}\color[HTML]{FFFFFF}14 & \cellcolor{heatmap15}\color[HTML]{FFFFFF}15 \\
				\hline
			\end{tabular}%
		}
		\fi
	\end{center}
	\label{table:confusionmatrix}
\end{table}
\endgroup

In total, we have 32 positive, 32 negative, and 32 neutral statements. In case of the neutral statements, 29 of 32 have been identified as neutral by the participants of our study, but only 21 negative and 10 positive. In these cases, the median labels of our participants coincide with the predefined labels of the data set. The participants identified two neutral and one negative statements as positive. Sixteen positive and 10 negative statements have been identified as neutral. In addition, the participants rated six positive and one neutral statement as negative. 

The confusion matrix of the Stack Overflow data set in Table~\ref{table:confusionmatrix} shows that no true positives have been classified as negative from the median participant labels. The same is true for true negative and positive as well as true neutral and negative. In addition, we observe that only 6 statements (12.5\%) were labeled as positive, while 33 (68.75\%) were labeled as neutral, and 9 (18.75\%) were labeled as negative. For the negative sentiment polarity class, only 9 statements were labeled as negative, instead of the authors 16 statements.

For the GitHub data set, true neutral statements are never classified as positive by the median participant labels, each other sentiment polarity class has been misclassified as one of the other two polarity classes at least once. More so, there are as many correctly classified true positive statements as false negative ones. This is notably interesting as we selected 16 statements from each sentiment polarity class of author labels, making an even split for each sentiment class out of the total 48 statements from each data set. However, we observed a notable discrepancy in the distribution of the median sentiment polarity classes calculated from the participant perceptions. 
Similarly, we observe that only 7 statements have been labeled as positive, while 22 were labeled as neutral, and 19 were labeled as negative. In both cases, the positive sentiment polarity class was assigned notably less then in the pre-labeled data set (6 and 7 times instead of 16). On the other hand, more statements (19 instead of 16) have been annotated as negative than by the pre-labels.

In both cases, the neutral label was chosen frequently by the participants, given that for the GitHub data set 22 instead of 16 statements were annotated with neutral, and for the Stack Overflow data set more than twice (33 instead of 16) the number of statements were annotated with neutral compared to the predefined labels, making up more than $\frac{2}{3}$ of the overall annotated sentiment labels.

\begin{finding}
    \textbf{Finding 6:} Comparing the median participants' label and the predefined labels, we observe some kind of pessimism by the participants. They rarely assign the positive class to a statement (in particular compared to the negative class).\\
    \textbf{Finding 7:} Considering the confusion matrices, we again observe a better performance of the guideline-based data set compared to the ad hoc labeled one. 
\end{finding}

\subsubsection{Cases of Severe Disagreement} 
As depicted in Table~\ref{table:rq1.1} and Table~\ref{table:confusionmatrix}, we observe 7 cases of severe disagreement between the calculated median labels of our participant's and the original labels provided by the authors for the guideline based GitHub data set. Since these cases are arguably the most interesting ones, we investigated the 7 cases manually to try to derive reasons for the statements' controversial nature. Out of the 7 cases of severe disagreement only one statement was labeled negative by the authors and perceived as positive by the participants, for the other six cases the perception was vice versa.

\begin{description}[style=unboxed,leftmargin=0cm]
\item\textbf{Statement:} \textit{``oh nice find, that's been bugging the crap out of me''}
\item\textit{Possible Explanation:}
We assume that this statement was labeled as negative by the authors because it contains the words \textit{``bugging''} and \textit{``crap''}, which indeed, on their own, are afflicted with a negative sentiment polarity. However, they only occur in the second half of a compound sentence. In the first half, the sentence has a positive sentiment polarity, praising the other person (presumably for finding a bug in the software), and only then explaining his appraisal with the explanation that what the other person has solved annoyed him/her beforehand.
\end{description}

One of the opposing example where the author label was positive and the calculated median label of the participants was negative was the following:

\begin{description}[style=unboxed,leftmargin=0cm]
\item\textbf{Statement:} \textit{``I have no idea either, I just trust the spray guys''}
\item\textit{Possible Explanation:}
We assume the authors have chosen the label positive because the word \textit{``trust''} was used, which has a positive sentiment polarity. The term \textit{``trust''} is also mentioned in the emotion framework by Shaver et al.~\cite{shaver1987emotion} under the category ``Love'' that was used by the authors: ``\dots the raters were trained to provide a polarity label based on the emotion detected according to the Shaver model\dots''~\cite{novielli2020githubgold}.  The term \textit{spray} used here is slang and means ``\textit{To use an automatic weapon to fire blindly and rapidly, releasing a large amount of bullets at one time.}'', according to \textit{Urban Dictionary}. In this sentence the term can be interpreted so that the writer has no idea to why a solution to a bug works, he just trusts randomly attacking the issue with common methods. This is probably a reply to a question as to why the solution given prior fixes the problem. Since the writer has no helpful explanation himself however, the statement can be perceived as negative, which was the case for the participants. Indeed the criterion of helpfulness was also mentioned by the participants as being used for annotating the sentiment polarities, which we discuss in more detail in subsection~\ref{subsec:criteria}.
\end{description}

Overall we observe, that in the cases of severe disagreement in the guideline based GitHub data set, the author labels usually reflect the sentiment polarity emitted by individual words that compose only parts of a compound sentence. The participants on the other hand tried to guess the context of the message reflected in the information given by the statement. In the given examples above we demonstrated, that this can make a significant difference. The annotation criteria used by the participants is described in subsection~\ref{subsec:criteria} and supports these findings further.


\subsection{Match of Participant Perceptions}
In a next step, we compared every single participant's selection of polarity classes with the predefined labels of the data set. 

\subsubsection{Agreement}
\begin{figure}[htbp]
	\begin{center}
		\resizebox{0.6\columnwidth}{!}{\input{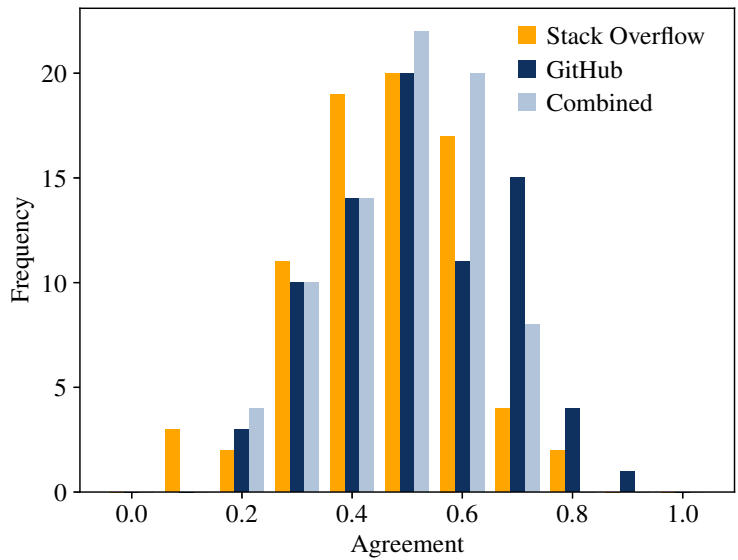}}
	\end{center}
	\caption{Frequency distribution of agreement ($n = 78$)}
	\label{fig:comparison_acc}
\end{figure}

Table~\ref{table:comparison_acc} summarizes the agreement of the participant's labels in comparison to the predefined labels. For the combined data set, we have a maximum agreement of 0.75 and a minimum  of 0.167 (mean: 0.489, SD: 0.137). For the Stack Overflow data set, we have a minimum of just 0.091 and a maximum of 0.833 (mean: 0.47, SD: 0.147). The GitHub data set performs slightly better than Stack Overflow with a minimum of 0.188, a maximum of 0.917, and a mean agreement of 0.517 (SD: 0.158). The distributions of all participants' agreement values for the Stack Overflow and GitHub data set are shown in Figure~\ref{fig:comparison_acc}. 


\begin{table}[htbp]
	\caption{Distribution of the calculated agreement values}
	\begin{center}
		\begin{tabular}{@{}llllll@{}}
		    \Xhline{2\arrayrulewidth}
			Data set & min & mean & max & SD & $n$ \\
			\hline
			Combined & 0.167 & 0.489 & 0.750 & 0.137 & 80\\  
			\hline
			Stack Overflow & 0.091 & 0.470 & 0.833 & 0.147 & 83\\
			\hline
			GitHub & 0.188 & 0.517 & 0.917 & 0.158 & 82\\
			\Xhline{2\arrayrulewidth}
		\end{tabular}
		\label{table:comparison_acc}
	\end{center}
\end{table}

\subsubsection{Cohen's Kappa Between the Participants and the Predefined Labels}

Table~\ref{table:comparison_kappa} summarizes the agreement between each participant and the predefined labels based on Cohen's $\kappa$. A negative $\kappa$-value indicates an agreement lower than by chance ~\cite{landis1977measurement}. 

For the combined data set, the $\kappa$-values range from -0.253 to 0.627 (mean: 0.235, SD: 0.203). 

For the Stack Overflow data set, the $\kappa$-values range from -0.314 to 0.75 with a mean of 0.208, which is considered slight agreement~\cite{landis1977measurement}, and a standard deviation of 0.208 ($n = 83$). Remarkably, 75\% of the $\kappa$-values are below 0.354, which is only considered fair agreement (50\% = 0.219, 25\% = 0.093). As with the agreement before, the GitHub data set performs better in terms of Cohen's $\kappa$. 

For the GitHub data set, the participants' $\kappa$-values range from -0.219 to 0.875, with an average of 0.275 and a standard deviation of 0.238. The 75\% quartile is at 0.5, which is considered moderate agreement~\cite{landis1977measurement}. 

\begin{table}[htbp]
	\caption{Distribution of the calculated $\kappa$-values}
	\begin{center}
		\begin{tabular}{@{}llllll@{}}
		    \Xhline{2\arrayrulewidth}
			Data set & min & mean & max & SD & $n$ \\
			\hline
			Combined & -0.253 & 0.235 & 0.627 & {0.203} & 80\\  
			\hline
			Stack Overflow & -0.341 & 0.208 & 0.750 & 0.208 & 83\\
			\hline
			GitHub & {-0.219} & {0.275} & {0.875} & 0.238 & 82\\
			\Xhline{2\arrayrulewidth}
		\end{tabular}
		\label{table:comparison_kappa}
	\end{center}
\end{table}

\begin{figure}[htbp]
	\begin{center}
		\resizebox{0.6\columnwidth}{!}{\input{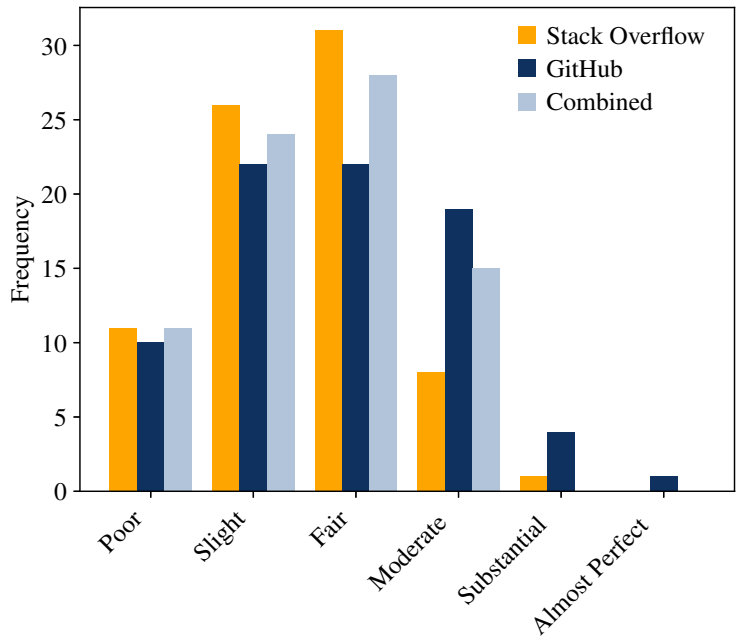}}
	\end{center}
	\caption{Frequency distribution of agreement strength classes, according to Landis and Koch~\cite{landis1977measurement} ($n = 78$)}
	\label{fig:comparison_kappa}
\end{figure}

Figure~\ref{fig:comparison_kappa} shows the distribution of the calculated agreement (by means of Cohen's $\kappa$) between each participant and the predefined labels for both data sets (divided into the classes proposed by Landis and Koch~\cite{landis1977measurement} to interpret the $\kappa$-values). Seemingly, the guideline annotated GitHub data set performs slightly better than the ad hoc labeled Stack Overflow data set.


However, to analyze this difference quantitatively, we tested the hypotheses presented in Table~\ref{table:hypotheses}. The Shapiro-Wilk test for normal distribution revealed that the agreement and Cohen's $\kappa$ are normally distributed for both data sets (cf. Table~\ref{table:norm}). Therefore, we used the repeated-measures t-test for H1.1 and H1.2.

\begin{table}[ht]
	\caption{Results of the Shapiro-Wilk test for normal distribution}
	\begin{center}
		\begin{tabular}{@{}lllll@{}}
		    \Xhline{2\arrayrulewidth}
			Data Set & Variable ($X$) & $W$ & $p$ & $X \sim N$\\
			\hline
			Stack Overflow & Agreement & 0.985 & 0.499 & Yes\\
			GitHub & Agreement & 0.985 & 0.496 & Yes\\
			\hline
			Stack Overflow & Cohen's $\kappa$ & 0.990 & 0.778 & Yes\\
			GitHub & Cohen's $\kappa$ & 0.985 & 0.504 & Yes\\
			\Xhline{2\arrayrulewidth}
		\end{tabular}
		\label{table:norm}
	\end{center}
\end{table}

The results of the repeated-measures t-test are shown in Table~\ref{table:ttest}, revealing significant differences for both data sets by means of agreement and Cohen's $\kappa$. As both $p$-values are below the significance level of $0.05$, we can reject H1.1$_0$ ($T$ = -3.395, $p$ = 0.00109 $<$ 0.05) and H1.2$_0$ ($T$ = -3.56, $p$ = 0.00063 $<$ 0.05). Summarizing, we can also reject H1$_0$, as the $p$-values are below the corrected significance level, and, thus, we can assume that there is a difference between the two data sets with regard to the match between the predefined and the participants' labels.  

\begin{table}[ht]
	\caption{Results of the repeated-measures t-test}
	\begin{center}
		\begin{tabular}{@{}llll@{}}
    		\Xhline{2\arrayrulewidth}
			Variable & $T$ & $p$ & $p < p_{corr}$\\
			\hline Agreement & -3.395 & 0.00109 & Yes\\
			Cohen's $\kappa$ & -3.560 & 0.00063 & Yes\\
			\Xhline{2\arrayrulewidth}
		\end{tabular}
		\label{table:ttest}
	\end{center}
\end{table}



\begin{finding}
    \textbf{Finding 8}: There is a significant difference in the agreement of the participants with the predefined labels comparing the guideline-based and the ad hoc labeled data sets.\\
    \textbf{Finding 9}: There is a significant difference in Cohen's $\kappa$ of the participants with the predefined labels comparing the guideline-based and the ad hoc labeled data sets.
\end{finding}




\subsection{Annotation Criteria}
\label{subsec:criteria}
After labeling the statements, we asked the participants what criteria they used for annotating the statements. Predefined answers included the tone (i.e. the mood implied by the writer's choice of words and the emotions they invoke on the reader), the content (i.e. the information of the statement, is something good or bad described?), and other (text field); multiple answers were possible. We placed this question after the labeling process on purpose, so that participants would label the statements without any influence, where as placing this question beforehand could lead to thinking about the options and choosing one beforehand and sticking with that choice. Out of 62 total responses to this question, 57 participants (91.9\%) stated that they used the tone of the statements, while 43 (69.4\%) stated that they used the content, and 15 (24.2\%) entered custom answers (other) in addition. We evaluated the 15 custom participant answers to this question by hand for a further analysis. Out of the 15 custom answers we found that 6 participants mentioned the use of emoticons (e.g., ``:-)''), or other slang (e.g., ``LOL'') for determining the sentiment polarity. The constructiveness of a post by being helpful, informative, or giving a solution to a problem was considered by 5 participants. Another 4 participants stated some form of guessing the context of the statement and guessing their own emotional response to it, based on their experience (e.g., ``Basically, I tried to guess the tone of the message, reflecting how the person typing the comment or reading it might feel,  as opposed to just communicating a technical fact.''). The lack of context was also criticized by one participant as making the annotation task difficult. Despite this fact, the context is still not considered by the sentiment analysis tools.

\section{Discussion}
\label{sec:discussion}
We end this paper by discussing our results, answering the research questions, and presenting threats to validity. 

\subsection{Answer to Research Questions}
Based on our results, we can answer the research questions as follows:

\textbf{RQ1:} We observe a huge difference between the median labels of the study participants and the predefined labels. From the 96 statements, the median label of the participants only coincides with the predefined labels in 60 cases, leading to an agreement of 0.625. 

\textbf{RQ2:} Looking at every single participant's perception in comparison to the predefined labels, we again observe a wide variety in the agreement. There are participants who achieve a very high agreement (for the combined data set, the maximum is at 0.75), but others achieve very low values, pointing to substantial disagreement by means of Cohen's $\kappa$. 

\textbf{RQ3:} In almost all cases, the participant's labels coincide more with the labels predefined by the GitHub data set than the Stack Overflow data set. That is, a data set that is labeled using concrete guidelines seems to better reflect the average perception of software project team members than an ad hoc labeled data set. The statistical tests also confirm that there is a significant difference between the two data sets with respect to agreement and Cohen's $\kappa$. However, we also observed a non negligible amount of severe disagreements between the participant's and authors in case of the guideline-based GitHub dataset, while the participant's only had mild disagreements with the original labels from the ad hoc labeled Stack Overflow data set.

\subsection{Interpretation}
\label{sec:interpretation}

Based on the results of our study and the answers to our research questions, we make three remarkable observations:

\begin{itemize}
    \item[(1)] Although the median values coincide with the predefined labels of the data sets in 62.5\% of the cases, we observe a huge difference between the single participant's ratings and the labels.
    \item[(2)] There is not a single participant who totally agrees with the predefined labels; ranging from barely substantial agreement to substantial disagreement.
    \item[(3)] In most cases, the labels of the guideline-based data set coincide more with the median participant(s) labels than the labels of the ad hoc labeled data set.
\end{itemize}

Besides the threats to validity that may have influenced our results (see Section~\ref{subsec:threats}), there are other possible explanations for these results. 

(1) Both data sets integrated in our study are meant to serve as a training set for sentiment analysis tools. That is, the labels assigned to the statements in the data set shall represent the perception of a ``typical'', i.e., of an ordinary member of a development team. Thus, the observation that the median perception of the participants coincides with the predefined labels in the data sets would have been expected. 

(2) However, despite the fact that the participants agree on average with the predefined labels, we observe remarkable discrepancies between single participants and the predefined labels. As described in Section \ref{sec:background}, it has already been mentioned by Imtiaz et al. \cite{imtiaz18sentiment} that label assignment by human raters without a coding scheme could lead to different understanding of sentiments in the field of software engineering among them. Novielli et al. \cite{novielli2020githubgold,N.Novielli.2018benchmark} showed that the absence of clear guidelines for annotation can lead to noisy gold standards data sets. However, our results showed a similar discrepancy between ad hoc labels assigned by our participants compared with both ad hoc labeled and guideline-based labeled data sets.

So apart from the explanations just mentioned, there are two other possible explanations (next to the threats to validity) for this observation. On the one hand, these discrepancies can emerge from the actual mood of the participants. Probably, in a software project setting, they may perceive single messages differently compared to this study setting. In a work situation, i.e., in a professional context, some statements might raise different feelings. On the other hand, the observation may be due to a general problem of sentiment analysis that has, to the best of our knowledge, not yet been addressed: Even though the predefined labels coincide with the median perception of a group of computer scientists, e.g., a development team, this is not necessarily true for single persons. In particular, as this kind of data base is used for the training of sentiment analysis tools, these tools also only reflect the median mood of the target group. That is, if a sentiment analysis tool is used in a team that tends towards very good or very bad mood (i.e., that appears to be optimistic rather than pessimistic, or vice versa) it is likely that the tool does not adequately reflect the real sentiment in this particular team. Therefore, it is worth a thought to focus on calibrating sentiment analysis tools to specific teams in future research. This would help increasing the accuracy of the analyses, which would in turn increase the trust of the team in the results. 

(3) Regarding the ad hoc labels of the participants, according to our results, the guideline-based data set performs better in almost all cases compared to the ad hoc labeled data set. This is somewhat surprising, as higher agreement between ad hoc labels (of the participants) and ad hoc labels (in the data set) was expected than between ad hoc and guidelines-based labels. The main difference between these two data sets is the way, the raters are asked to assign the labels to the statements. In case of the ad hoc labeling, the raters are asked to label the statements according to their perception without further thinking about it. This leads to some kind of unorganized rating process, but this way of labeling coincides with ``reality''. A receiver of a message will unlikely think about how to interpret such statement (i.e., whether it is meant to be positive, negative, or neutral), but will trust his or her gut feeling. However, the guideline-based labeling process, in which the raters apply specific guidelines when assigning the polarity class to a statement, appears to be more stringent with the gut feeling of the participants in our study, as we did not tell them how to label the statements. Most of them used the tone and the content, which raise an impression by the reader, and thus reflect the gut feeling. 

As Imtiaz et al. \cite{imtiaz18sentiment} have also pointed out, there seem to be different reasons for assigning a sentiment. For example, participants might assign labels based on the perception of themselves being the receiver of a message, or they might put themselves in the position of the sender. They could also label sentiments based on the content. 
Novielli et al. \cite{novielli2018stackgold} did not have high Fleiss' $\kappa$ values in their emotion attribution. Their focus was on emotion recognition to build their data set. But it is debatable whether this emotion aspect is sufficient for a complete sentiment analysis, or if you still have to look at the other aspects mentioned before. Our results show that computer scientists do not seem to focus only on emotions in their perception. Or their perception of emotion differs from that of the authors. This means that tools trained with such data sets are limited in their practical applicability.

\subsection{Implications}

Independent of the selected labeling process, our results point to the necessity of an increased awareness of the data sets to be used when training sentiment analysis tools, as this selection strongly influences the outcome of a tool. Thus, the data base should be selected with care and should be appropriate for the pursued use case. But also after having selected an appropriate data set, it is worth a thought whether the tool still needs to be adjusted to the given context, including the team in which it should be used. This could be done, for example, by randomly examining the communication data manually and calculating the balance of negative, neutral, and positive, and matching it with that of the data set.  However, this requires further research. Another option that, however, requires further research would be some kind of calibration to the team with its specific characteristics. Do we use the tool in a team that generally has a very good mood or is it a rather depressed team? Is there a lot of irony in the team communication that in addition increases the risk of misclassifications? As all these questions have been sparsely, if at all, considered in literature, further studies are required to explain how this adjustment can be achieved. 

In addition, when developing sentiment analysis tools, researchers should ask themselves what they want to predict. We do not claim that the results of our study are always correct, but our findings highlight that the predefined labels of the data sets are not correct for each and every person (what the authors of the data sets do not claim). However, the accuracy of sentiment analysis tools is often tested against these predefined labels. This raises the question how meaningful such kind of validation is if the labels of the data sets do not coincide with the perceptions of the team. The gold standard data sets are a good option to test the general accuracy and performance of sentiment analysis tools, but it does not allow profound conclusions on the applicability in a specific team. And this is what researchers and practitioners should keep in mind when using such tools. The tool can always only be as good as the underlying data, but it is questionable how well the underlying data fits the team. 

Summarizing, the results of our study highlight the need for awareness when applying sentiment analysis tools such that they fit the given context and the team. That is, both the origin of the training set and, hence, the application domain of the sentiment analysis tool, and the subjective labels assigned to the training data must fit the perceptions of the team. In addition, as sentiment analysis reflects subjective perceptions, results should be handled with care. The tools that produce the results are trained on subjective data (that might be made kind of objective by referring to guidelines) and, thus, the outcome should be seen as an indicator rather than the ultimate truth.

\subsection{Threats to Validity}
\label{subsec:threats}
Our results are limited to our (selected) survey population, and cannot be generalized for all developers or computer scientists. In this section, we summarize the most relevant threats to validity possibly impacting our results.

Due to the location of the researchers, the vast majority of our population were non-native English speaker. Nevertheless, the statements from the data sets were entirely in English including many technical terms. To countermeasure this threat, we asked our participants about their frequency of communication in English. About two thirds of our population reported to communicate in English once a week or more, while one third communicated in English once in a while or less. Nevertheless, we assume that all participants considered their own English comprehension suitable for performing the survey. In addition, almost all technical terms from the software engineering or computer science domain are English independent of the language used.

Due to the nature of a survey study, the participants answered the survey questions at home, leading to possible distractions and interferences for individual participants.

The study was also distributed among colleagues with the request to distribute the study to other potential candidates. Therefore, it is possible that raters of the two data sets used in our paper also participated. We could not exclude this in advance, but we found it negligible due to the high number of participants.

The vast majority of over two thirds of our population were students in the computer science field, and not professional developers. We consider this threat negligible, since even gold standard data sets are often created with the annotations of computer science students (cf.~\cite{novielli2018stackgold}), and not long-standing professional developers. However, we will investigate our gathered data on group specific behavior in future research. For all that, we only included participants who identified as (prospective) computer scientists or had programming experience. We are confident that the median labels of our participants reflect the perception of an average individual in the software engineering domain.

The capability of participants to annotate a sentiment polarity to a message can be aggravated by the lack of the messages context. We only presented single, randomly selected messages from the original data sets and no course of a conversation. Some participants mentioned trying to guess the context of a response in order to be able to choose a proper sentiment polarity class for it. Although we are aware of this, we wanted to gather the initial emotional responses for the statements instead of presenting the participants with an annotation criteria beforehand and making them act accordingly when selecting the sentiment polarity classes.

\section{Conclusion}
\label{sec:conclusions}
Sentiment analysis tools strongly rely on pre-labeled data sets that provide polarity classes for specific statements. These polarity classes shall reflect the perception of a somewhat ``typical'' developer as the resulting tools are meant to be able to predict his or her perception. 

In order to compare the perceptions of potential software project team members with the predefined labels, we conducted an online survey. Based on 94 data points, we compared the median perception of the participants with the predefined labels, leading to a match between the perceptions and the labels in 62.5\% of the cases.

In a next step, we concentrated on single participants and evaluated their agreement with the predefined labels, leading to results ranging from systematic disagreement to almost perfect agreement. This points to a potential need to adjust the tools to the personalities of a team, which should be addressed in future research.

On average, the agreement between single participants and the group of participants is better in case of the guideline-based labeled data set compared to the ad hoc labeled one, despite the fact that the participants labeled the statements also according to their gut feeling (without any guidelines) in our survey. 

These results should increase the awareness of the need to carefully select the training data set for the development of sentiment analysis tools, and to handle the results with care as perceptions are very subjective and the forecast of a sentiment analysis tool should not be over-interpreted. Summarized, the results of sentiment analysis in project contexts should be taken as a grain of salt rater than as the ultimate truth.

\section*{Acknowledgment}
This research was funded by the Leibniz University Hannover as a Leibniz Young Investigator Grant (Project \textit{ComContA}, Project Number \textit{85430128}, 2020--2022).


\bibliographystyle{elsarticle-num}
\bibliography{references.bib}

\end{document}